\documentclass[reprint, aps,prl,nofootinbib,floatfix]{revtex4-1}
\usepackage[utf8]{inputenc}
\usepackage{amsmath, amsthm, amssymb, amsfonts, bbm, mathrsfs}
\usepackage{graphicx}
\usepackage{float}
\newcommand{\bra}[1]{\langle #1 \vert}
\newcommand{\ket}[1]{\vert #1 \rangle}
\newcommand{\inner}[2]{\langle #1 \vert #2 \rangle}

\begin{document}

\title{Probing Quantum Dynamical Couple Correlations with Time-Domain Interferometry}

\author{Salvatore Castrignano}
\affiliation{Max-Planck-Institut f\"ur Kernphysik, Saupfercheckweg 1, 69117 Heidelberg, Germany}

\author{J\"org Evers}
\affiliation{Max-Planck-Institut f\"ur Kernphysik, Saupfercheckweg 1, 69117 Heidelberg, Germany}

\begin{abstract}
Time domain interferometry is a promising method to characterizes spatial and temporal correlations at x-ray energies, via the so-called intermediate scattering function and the related dynamical couple correlations. However, so far, it has only been analyzed for classical target systems. Here, we provide a quantum analysis, and suggest a scheme which allows to access quantum dynamical correlations. We further show how TDI can be used to exclude classical models for the target dynamics, and illustrate our results using a single particle in a double well potential.
\end{abstract}

\maketitle

\textit{Introduction} ---
Spatial and temporal correlations among particles are key to the exploration of complex many-body phenomena. Scattering experiments provide access to the scattering function $S(\mathbf{p},\omega)$ that is proportional to the cross section for scattering with energy transfer $\hbar \omega$ and momentum transfer $\hbar \mathbf p$~\cite{VANHOVE}. It characterizes the evolution of correlations on time scales $\sim 1/\omega$ and length scales $\sim 1/|\mathbf{p}|$. In practice, knowledge of the correlations over a broad range of time and momentum transfer scales is desirable, and various scattering techniques such as x-ray~\cite{XFEL,SWISSFEL2017} and neutron~\cite{LOVESEY} scattering can be used to access complementary  energy and momentum transfer scales. Similarly, depending on the properties of the scatterer, it can be favorable to characterize correlations directly in the time domain, via the intermediate scattering function (ISF)
\begin{align}
S(\mathbf{p},t_1,t_2)=\int_V G(\mathbf{r},t_1,t_2) e^{i \mathbf{p}\cdot\mathbf{r}} d^3 r\,,
\end{align}
with the dynamical couple-correlation function (DCF) 
\begin{align}
\label{eqn: DCF}
G(\mathbf{r},t_1,t_2)=\int_V \text{Tr} \big[ \mu \, \hat{\rho}(\mathbf{r}',t_1) \hat{\rho}(\mathbf{r}'+\mathbf{r},t_2) \big] d^3 r'\,.
\end{align}
Here, the system described by the density matrix $\mu$ covers the volume $V$, and $\hat{\rho}(\mathbf{r},t)$  is the particle-density operator. The DCF quantifies the spatial and temporal correlations between particles at $(\mathbf{r}, t_1)$ and   $(\mathbf{r}'+\mathbf{r}, t_2)$.

\begin{figure*}[t]
  \centering
    \includegraphics[width=0.9\textwidth]{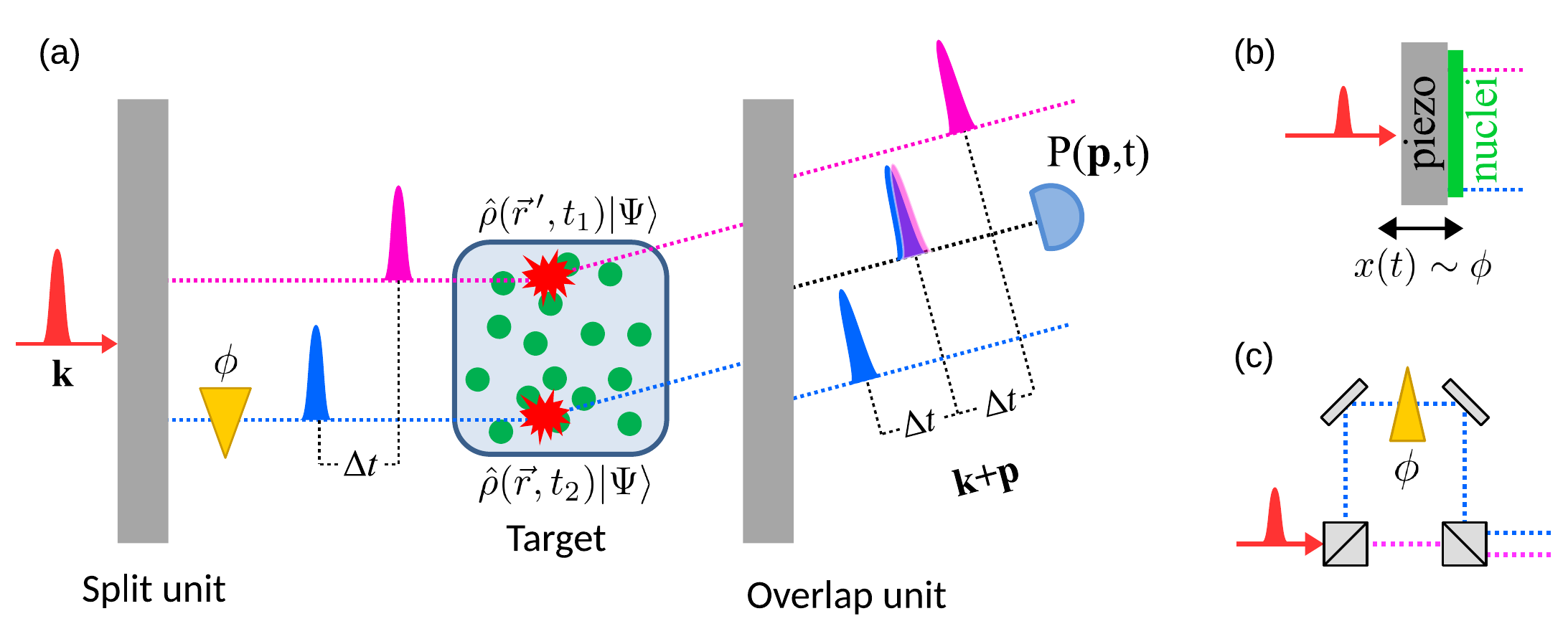}
    \caption{(Color online) (a) Schematic setup. The incoming wave packet propagating along $\mathbf{k}$ (red) is separated into two parts with a mutual delay by a ``split unit''. The advanced (violet) component is quasi-elastically scattered by the target at time $t_1$, while the delayed one (blue) scatters at $t_2$. Subsequently, the light scattered in direction $\mathbf{k}+\mathbf{p}$ passes an ``overlap unit'', which acts identical to the split unit. The central component of the outgoing signal contains two indistinguishable contributions, arising from photons which scattered at time $t_1$ or $t_2$, respectively. In our scheme, a phase shifter $\phi$ controls the interference of these two contributions in the measured intensity of the scattered light, and thus enables one to recover the quantum dynamical couple correlation function of the target.  (b) If the split and overlap units are realized using M\"ossbauer filter foils, then the required phase control is possible using mechanical displacements of the split foil as demonstrated in~\cite{Heeg2017}. (c) A generic implementation of the phase control is a split-and-delay line, with a phase plate in one of the two arms.  }
\label{fig: TDIScheme}
\end{figure*}

A particular technique to access the ISF is the so-called time-domain interferometry (TDI)~ \citep{BARON,SMIRNOV,SMIRNOV2006,Saito2012,Saito2017,Kaisermayr2001,1882-0786-2-2-026502} (see Fig.~\ref{fig: TDIScheme} for the extended scheme used here). It has recently been suggested as a promising candidate for x-ray free electron laser experiments (see page 84 of~\cite{XFEL}; note that the general feasibility of free-electron-laser experiments with M\"ossbauer nuclei has already been demonstrated in a different setting~\cite{chumakov}). TDI allows to measure ISF over much longer times than competing techniques, and since it is essentially background-free even for  intense x-ray pulses. TDI uses filter foils containing long-lived M\"ossbauer isotopes, which are placed in front of and behind the actual target. The incident x-ray frequency is chosen in resonance with the M\"ossbauer nuclear transition. The first foil (which pictorially can be thought of as a ``split unit'') induces  two possible scattering channels for the incoming pulse. The first prompt channel comprises photons which did not interact with the nuclei. The photons in the second channel are delayed in time, due to the interaction with the long-lived nuclear transition. As a consequence, the photons in the two channels probe the target at different times $t_1, t_2$. After the interaction, the second M\"ossbauer foil (``overlap unit'') again splits each of the two channels into a prompt and a delayed contribution. This ``overlap operation'' creates scattering channels to the detected signal, which were either delayed in the split unit or in the overlap unit, but not in both, and thus reach the detector at the same time. For these, it is not possible to distinguish if the interaction with the target took place at time $t_1$ or $t_2$, and the interference of these two pathways leads to temporal modulations of the detection signal, which in turn provide access to the ISF. Depending on the chosen M\"ossbauer species, different momentum and energy transfer ranges can be accessed~\cite{1882-0786-2-2-026502}. Recently, also a modified scheme using M\"ossbauer foils with two resonances has been suggested~\cite{doi:10.1063/1.5008868}. 

So far, TDI has been analyzed and demonstrated experimentally~\cite{BARON,Saito2012bis,Saito2014,Kaisermayr2001,1882-0786-2-2-026502,doi:10.1063/1.5008868} for targets which can be described by classical mechanics~\cite{HANSEN}. However, quantum effects change the DCF~\cite{VANHOVE,VANHOVE1}, and such quantum corrections have been theoretically studied~\cite{Watson,barocchi,sears,schofield} and observed in quantum-liquids~\cite{cunsolo05,cunsolo03} or in surface diffusion~\cite{surface1}. In thermal equilibrium,  quantum effects are usually considered to be restricted to relatively short times of order $\hbar/(k_B\,T)$~\cite{schofield}. One obvious solution is to lower the temperature, which is taken to the extreme in cold-gas implementations of solid-state dynamics~\cite{Gross995,RevModPhys.86.153}, where quantum effects were observed in the response functions using inelastic light scattering~\cite{landig}. But more importantly, a central research goal of modern x-ray sources is the study of strongly correlated and quantum materials, in and out-of equilibrium. Their features largely depend on quantum phenomena (see, e.g., ~\cite{keimer,Dagotto257,QUINTANILLA}), and they exhibit correlations over a broad range of temporal and spatial scales, in particular out-of-equilibrium.  In this regard, the  nano­second to millisecond scale is considered very interesting, but hard to access experimentally~\cite{petra4,xpcs}. This raises the question, whether time domain techniques can be used to explore correlations in targets which require a quantum mechanical treatment,  this implying a modification of their DCF and ISF by the above mentioned quantum corrections.

Here we provide a quantum mechanical analysis of TDI, and suggest a scheme which allows one to measure the ISF both for quantum and classical targets. In our scheme, the full ISF is accessed by controling the interference between the different scattering channels via their relative phase. DCF and ISF have different properties for classical and quantum targets and we show how TDI can be used to exclude classical models for the targets.  Finally, we illustrate our main results with a minimal model composed of a single particle hopping between two sites.

\textit{Properties of DCF and ISF} --- We start with symmetry properties of DCF and ISF, which will enable us to distinguish quantum mechanical targets from classical ones. As already noted by van Hove himself in \cite{VANHOVE,VANHOVE1}, for quantum systems (subscript $qu$), the DCF is in general a complex-valued function due to the non-commutativity of particle-density operators at different times.
It directly follows from definition Eq.~(\ref{eqn: DCF}) that
\begin{align}
G_{qu}(\mathbf{r},t_1,t_2)^* &= G_{qu}(-\mathbf{r},t_2,t_1)\,, \label{eqn: SymmDCF}\\
S_{qu}(\mathbf{p},t_1,t_2)^* &= S_{qu}(\mathbf{p},t_2,t_1)\,. \label{eqn: SymmISF}
\end{align}
If the system instead is described by a classical model (subscript $cl$), the density of particles is a real valued function, and  the quantum mechanical trace is replaced by a statistical ensemble average in Eq.~(\ref{eqn: DCF}). As a consequence, the classical DCF is a real-valued function, giving rise to a different behavior of the ISF under complex conjugation,
\begin{align}
\label{eqn: SymmClassISF}
S_{cl}(\mathbf{p},t_1,t_2)^*=S_{cl}(-\mathbf{p},t_1,t_2)\,.
\end{align}
Note that not only the sign of $\mathbf{p}$ is changed as compared to the quantum case Eq.~(\ref{eqn: SymmISF}), but also the order of the time arguments $t_1, t_2$.

\textit{Quantum theory of TDI} ---
We now turn to the analysis of TDI in the case of a quantum target (see Fig.~\ref{fig: TDIScheme}). In addition to the original TDI proposal, we assume that the relative phase $\phi$ between the scattering channels can be controlled. As we will show below, this enables control of the interference between the different scattering channels, and thereby provides access to the full ISF. For M\"ossbauer foils, the required phase control is possible with sub-\AA ngstrom precision on a nanosecond scale using mechanical displacements of the split foil, as demonstrated in~\cite{Heeg2017} (see Fig. 1 (b)). Related precise control of mechanical motion has also been demonstrated in~\cite{Vagizov2014}.
In order to simplify the discussion, we consider a setup in which the split and the overlap units separate incoming pulses into two identical copies with mutual delay $\Delta t$. One possible realization for this is a split-and-delay line with  a phase plate, see Fig. 1 (c). 
Behind the overlap unit, the signal is temporally separated into three pulses. The leading [trailing] pulse comprises those photons which were delayed in none [both] of the split and overlap units, and which interacted at time $t_1$ [$t_2$] with the target. In contrast, the central pulse contains photons which were either delayed in the split unit or in the overlap unit, but not in both. It is therefore not possible to distinguish if the interaction with the target took place at time $t_1$ or $t_2$. In the following, we will concentrate on this part. Note that the corresponding quantum analysis of the original setup with M\"ossbauer foils is given in the Supplemental Material~\cite{supplement}.

We proceed by calculating the probability amplitude that a photon from the central pulse is registered by a detector placed at position $\mathbf{R}$ at time $t$, by summing up the detection amplitudes for the two indistinguishable scattering pathways. These evaluate to ($j\in \{1,2\}$, see Supplemental Material~\cite{supplement} for details)
\begin{equation}
\label{eqn: DetectionAmpl}
\frac{e^{i \omega_0 (R/c - t )}}{R} e^{i\phi_j} f(t) \int_V d^3 r e^{-i\mathbf{p}\cdot\mathbf{r}}\hat{\rho}(\mathbf{r},t_j)\ket{\psi}\,.
\end{equation} 
As expected, the amplitudes are spherical wave packets with carrier frequency and envelope $f(t)$ identical to those of the incoming photon. The amplitudes depend on the target's density operator at the scattering times and on the initial state of the target $\ket{\psi}$. Here, $\mathbf{p}$ is the exchanged momentum between the photon and the target.
The signal recorded by the detector will be proportional to the probability of detecting the photon, which in turn is
\begin{align}
\label{eqn: DetectionProb}
&P(\mathbf{p},t) \propto  f(t)^2 \bigg( \sum_{j=1,2} S_{qu} (\mathbf{p},t_j,t_j) \nonumber \\
&+ 2 \cos[\phi]\,S^R_{qu} (\mathbf{p},t_1,t_2) - \sin[\phi]\,S^I_{qu} (\mathbf{p},t_1,t_2) \bigg)\,,
\end{align}
where $\phi=\phi_2-\phi_1$ is the phase difference between the two scattering pathways. Here and in the following, a superscript $R$ [$I$] denotes the real [imaginary] part, such that $S_{qu} = S_{qu}^R + i\, S_{qu}^I$. Note that Eq.~(\ref{eqn: DetectionProb}) applies to targets initially in a pure quantum state. Otherwise, it has to be averaged over  all possible initial states.

As our first main result, we find from Eq.~(\ref{eqn: DetectionProb}) that control over the relative phase $\phi$ and the delay $\Delta t$ enables one to individually access the real and the imaginary parts of the ISF as function of momentum transfer $\mathbf{p}$ and time $t$, as desired. 

Next, in order to extract information about the quantum or classical nature of the target, we consider the sum $I^+$ and the difference $I^-$ of the intensities at two opposite exchanged momenta $\pm \mathbf{p}$. Using Eq.~(\ref{eqn: DetectionProb}), 
\begin{align}
\label{iqu}
I_{qu}^{\pm}(\phi,t)  & \propto f(t)^2 \bigg( \sum_{j=1,2}  \big[ S_{qu}(\mathbf{p},t_j,t_j) \pm S_{qu}(-\mathbf{p},t_j,t_j) \big] \nonumber \\
+2 \bigg\{ \cos[\phi] & \big[ S^R_{qu}  (\mathbf{p},t_1,t_2) \pm  S^R_{qu} (-\mathbf{p},t_1,t_2) \big]+ \nonumber \\
- \sin[\phi] \big[  & S^I_{qu} (\mathbf{p},t_1,t_2)  \pm S^I_{qu} (-\mathbf{p},t_1,t_2) \big] \bigg\} \bigg)\,.
\end{align}
If a classical model for the target is assumed, such that the ISF satisfies the symmetry Eq.~(\ref{eqn: SymmClassISF}), then Eq.~(\ref{iqu}) simplifies to
\begin{align}
I_{cl}^{+}(\phi,t) \propto & f(t)^2 \bigg( \sum_{j=1,2} S(\mathbf{p},t_j,t_j) + \nonumber \\
+ & 2 \cos [\phi] \, S^R(\mathbf{p},t_1,t_2) \bigg)\,, \label{eqn: SymmetricalIpma}\\
I_{cl}^{-}(\phi,t) \propto & - 2 f(t)^2 \sin [\phi] \, S^I(\mathbf{p},t_1,t_2)\,.\label{eqn: SymmetricalIpmb}
\end{align}
Thus, recording $I_{\pm}$ for different values of $\phi$ enables  one to distinguish quantum or classical symmetries of the target. 
If, for example, $I_{-}$ does not vanish at $\phi=n\pi$, then the classical relation Eq.~(\ref{eqn: SymmetricalIpmb}) is ruled out. It follows that the ISF of the target has no inversion symmetry, such that the DCF is a complex valued function and a quantum model for the target is needed. 
In the opposite case,  DCF is real valued. Then, it may still be possible to violate Eq.~(\ref{eqn: SymmetricalIpma}) to exclude a classical model. However, it is important to note that a real DCF alone does not imply a classical target. Rather, also quantum targets may exhibit real valued DCF for particular parameter choices. This fact is explicitly shown for a concrete system in the next section.

\textit{Model} --- In the final part, we illustrate our results with a single particle in a double well potential. The DCF and ISF for this simple model can be calculated exactly, explicitly showing that a non-vanishing imaginary part of the DCF can be attributed to the existence of quantum coherences. These coherences arise, if the particle is in a coherent superposition of position eigenstates. However, the reverse is not true, since we find particular superposition states for which the DCF is real-valued.

We denote the two wells by $L$ and $R$, and the particle dynamics is governed by the Hamiltonian
\begin{align}
\label{eqn: DWHamiltonian}
H=-\hbar \, \frac{\Omega}{2} \; (\ket{L}\bra{R}+\ket{R}\bra{L})\,.
\end{align}
A generic state of the particle at time $t$ in the $\ket{L},\ket{R}$ representation is given by the density matrix
\begin{align}
\label{eqn: DensityMatrix}
\mu(t)=\begin{pmatrix}
P_L (t) & \Gamma (t)\\
\Gamma(t)^* & P_R (t)
\end{pmatrix}\,,
\end{align}
where $P_L (t), P_R (t)$ are the probabilities of finding the particle at time $t$ at position $j$ which satisfy the condition $P_L (t)+P_R (t)=1$, while $\Gamma (t)$ is the coherence coefficient. The DCF calculated for the state (\ref{eqn: DensityMatrix}) is
\begin{align}
G_{qu}(\mathbf{d},t_1,t_2) &=  \mathcal{S}^2 \, P_L(t_1) +\frac{i}{2}\mathcal{S}' \; \Gamma(t_1)^* \,, \label{eqn: DCFDoublewellD} \\
G_{qu}(-\mathbf{d},t_1,t_2) &=  \mathcal{S} ^2 \, P_R(t_1) +\frac{i}{2}\mathcal{S}'  \; \Gamma(t_1) \,,  \label{eqn: DCFDoublewell-D}\\
G_{qu}(0,t_1,t_2) &=  \mathcal{C} -i \mathcal{S}'  \; \Gamma^R(t_1)\,, \label{eqn: DCFDoublewell0}
\end{align}
where $\mathcal{S} = \sin[\Omega\, \Delta t/2]$,  $\mathcal{C} = \cos[\Omega \, \Delta t/2]$, $\mathcal{S}' = \sin[\Omega\, \Delta t]$, $\Delta t=t_2-t_1$, and $\Gamma^R(t_1)$ indicates the real part of $\Gamma(t_1)$. Expressions~(\ref{eqn: DCFDoublewellD})-(\ref{eqn: DCFDoublewell0}) are complex valued if $\Gamma^R(t_1)$ is non-zero, that is when the particle is in a
coherent superposition of $\ket{L}$ and $\ket{R}$. On the contrary, a
purely imaginary $\Gamma(t_1)$ gives a real DCF even though the state
is in a quantum superposition. Thus, we find that a real valued DCF alone does not imply classical behavior.

The ISF corresponding to~(\ref{eqn: DCFDoublewellD}-\ref{eqn: DCFDoublewell0}) is
\begin{equation}
\label{eqn: ISFDoublewell}
\begin{split}
S_{qu}(\mathbf{p},t_1,t_2) = \mathcal{S}^2 \cos & [\mathbf{p}\cdot\mathbf{d}] + \mathcal{C}^2 +\\
+i \bigg\{ \big[ P_L (t_1)-P_R (t_1) \big] & S^2 \sin[\mathbf{p}\cdot\mathbf{d}] + \\
+\frac{\mathcal{S}'}{2} \big[ \Gamma^I(t_1) \sin[\mathbf{p}\cdot\mathbf{d}] + \Gamma^R ( & t_1) \big( \cos[\mathbf{p}\cdot\mathbf{d}]-2 \big) \big] \bigg\}
\end{split}
\end{equation}
which evidently satisfies the identity~(\ref{eqn: SymmClassISF}) only if $\Gamma^R(t_1)=0$, consistent with the results for the DCF.
It turns out that $\Gamma^R$ is a constant of motion under the action of Hamiltonian Eq.~(\ref{eqn: DWHamiltonian}). This allows us to relate the results better to an actual experimental implementation, in which it may only be possible to control the delay $\Delta t$, but not $t_1$ itself. Averaging over $t_1$, we find
\begin{align}
\bar{G}_{qu}(\pm \mathbf{d},\Delta t) &=  \frac 12 \mathcal{S} + \frac i2  \mathcal{S}'\, \Gamma^R\,,\\
\bar{G}_{qu}(0,\Delta t) &= \mathcal{C} -i \mathcal{S}' \, \Gamma^R\,,\\
\bar{S}_{qu}(\mathbf{p},\Delta t) &= \mathcal{S}^2 \, \cos[\mathbf{p}\,\mathbf{d}] + \mathcal{C}^2\nonumber\\
& + \frac i2 (\cos[\mathbf{p}\,\mathbf{d}] - 2) \mathcal{S}' \, \Gamma^R\,.
\end{align}
As before, the complex nature of the DCF is linked to $\Gamma^R$. From Eq.~(\ref{eqn: SymmetricalIpmb}), we further find $\bar{I}^- = 0$, such that a classical model cannot be excluded. But $\bar{I}^+$ has a contribution proportional to $\Gamma^R\, \sin\phi$, which is at odds with Eq.~(\ref{eqn: SymmetricalIpma}) if $\Gamma^R\neq 0$, such that then a classical model can be excluded.

\textit{Summary and discussion} ---
DCF and ISF have different properties for quantum and classical systems. The non-commutativity of particle-density operators at different times in general leads to imaginary contributions to the DCF for quantum systems, and DCF and ISF have different symmetry properties under complex conjugation for classical and quantum systems. Using the quantum mechanical analysis presented here, we have shown that time-domain techniques can be used to measure the complex-valued ISF. Moreover, the comparison of the ISF at two opposite values of the exchanged momentum $\mathbf{p}$ in the form Eq.~(\ref{iqu}) provides access to the symmetry properties of the system's ISF,  and gives a handle to exclude classical models for the target. 
Throughout the analysis, we used a simplified model for the split and overlap units, but our results carry over to the case of M\"ossbauer filter foils (see supplemental materials~\cite{supplement}), for which the  required relative-phase control is possible with the necessary precision~\cite{Heeg2017,Vagizov2014}.

While quantum corrections to the DCF already appear in thermal equilibrium, a suitable preparation of the sample is expected to induce quantum effects, and to render them more accessible, e.g., by reducing detrimental averagings in the measurement. Pulsed laser systems synchronized to the x-rays are under development at most x-ray facilities, and have also already been demonstrated with M\"ossbauer nuclei~\cite{Sakshath2017}.
Our TDI scheme is not restricted to the x-ray domain, but could also be used to explore correlations on other time and length scales, such as cold-atom implementations of solid state dynamics~\cite{landig}. This requires the availability of suitable split and overlap units, and a system whose internal dynamics  has no resonance in the spectrum of the probing photon pulse, so that only quasi-elastic scattering of the photon is relevant. 
The analysis of our simple double-well model could serve as a starting point for the investigation of related phenomena in more realistic settings. For example, cold atoms trapped in atomic lattices serve as quantum simulators for complex solid state phenomena, structured periodic potentials appear on surfaces of materials, and certain complex materials may intrinsically offer various quantum states placed in a periodic potential landscape. Few-particle systems in single- or double-well potentials have also been studied directly~\cite{MURMANN}.
Finally, we note that the appearance of imaginary parts in such quantities poses practical and interpretative problems~\cite{BALLENTINE,MARGENAU,PhysRevA.96.022127}, which could be explored experimentally  using TDI techniques.

\begin{acknowledgments}
This work is part of and supported by the DFG Collaborative Research Centre ``SFB 1225 (ISOQUANT).''
\end{acknowledgments}

\bibliography{Bibliography}

\begin{thebibliography}{43}%
\makeatletter
\providecommand \@ifxundefined [1]{%
 \@ifx{#1\undefined}
}%
\providecommand \@ifnum [1]{%
 \ifnum #1\expandafter \@firstoftwo
 \else \expandafter \@secondoftwo
 \fi
}%
\providecommand \@ifx [1]{%
 \ifx #1\expandafter \@firstoftwo
 \else \expandafter \@secondoftwo
 \fi
}%
\providecommand \natexlab [1]{#1}%
\providecommand \enquote  [1]{``#1''}%
\providecommand \bibnamefont  [1]{#1}%
\providecommand \bibfnamefont [1]{#1}%
\providecommand \citenamefont [1]{#1}%
\providecommand \href@noop [0]{\@secondoftwo}%
\providecommand \href [0]{\begingroup \@sanitize@url \@href}%
\providecommand \@href[1]{\@@startlink{#1}\@@href}%
\providecommand \@@href[1]{\endgroup#1\@@endlink}%
\providecommand \@sanitize@url [0]{\catcode `\\12\catcode `\$12\catcode
  `\&12\catcode `\#12\catcode `\^12\catcode `\_12\catcode `\%12\relax}%
\providecommand \@@startlink[1]{}%
\providecommand \@@endlink[0]{}%
\providecommand \url  [0]{\begingroup\@sanitize@url \@url }%
\providecommand \@url [1]{\endgroup\@href {#1}{\urlprefix }}%
\providecommand \urlprefix  [0]{URL }%
\providecommand \Eprint [0]{\href }%
\providecommand \doibase [0]{http://dx.doi.org/}%
\providecommand \selectlanguage [0]{\@gobble}%
\providecommand \bibinfo  [0]{\@secondoftwo}%
\providecommand \bibfield  [0]{\@secondoftwo}%
\providecommand \translation [1]{[#1]}%
\providecommand \BibitemOpen [0]{}%
\providecommand \bibitemStop [0]{}%
\providecommand \bibitemNoStop [0]{.\EOS\space}%
\providecommand \EOS [0]{\spacefactor3000\relax}%
\providecommand \BibitemShut  [1]{\csname bibitem#1\endcsname}%
\let\auto@bib@innerbib\@empty
\bibitem [{\citenamefont {Van~Hove}(1954)}]{VANHOVE}%
  \BibitemOpen
  \bibfield  {author} {\bibinfo {author} {\bibfnamefont {L.}~\bibnamefont
  {Van~Hove}},\ }\href@noop {} {\bibfield  {journal} {\bibinfo  {journal}
  {Phys. Rev.}\ }\textbf {\bibinfo {volume} {95}},\ \bibinfo {pages} {249}
  (\bibinfo {year} {1954})}\BibitemShut {NoStop}%
\bibitem [{\citenamefont {Heyderman}\ \emph {et~al.}(2009)\citenamefont
  {Heyderman}, \citenamefont {Milne}, \citenamefont {Thibault}, \citenamefont
  {Ballmer},\ and\ \citenamefont {Staub}}]{XFEL}%
  \BibitemOpen
  \bibfield  {author} {\bibinfo {author} {\bibfnamefont {L.}~\bibnamefont
  {Heyderman}}, \bibinfo {author} {\bibfnamefont {C.}~\bibnamefont {Milne}},
  \bibinfo {author} {\bibfnamefont {P.}~\bibnamefont {Thibault}}, \bibinfo
  {author} {\bibfnamefont {K.}~\bibnamefont {Ballmer}}, \ and\ \bibinfo
  {author} {\bibfnamefont {U.}~\bibnamefont {Staub}},\ }\href@noop {} {\emph
  {\bibinfo {title} {Ultrafast Phenomena at the Nanoscale: Science
  opportunities at the SwissFEL X-ray Laser}}},\ \bibinfo {type} {Tech. Rep.}\
  (\bibinfo  {institution} {Paul Scherrer Institut},\ \bibinfo {year}
  {2009})\BibitemShut {NoStop}%
\bibitem [{\citenamefont {Abela}\ \emph {et~al.}()\citenamefont {Abela},
  \citenamefont {Beaud}, \citenamefont {van Bokhoven}, \citenamefont {Chergui},
  \citenamefont {Feurer}, \citenamefont {Haase}, \citenamefont {Ingold},
  \citenamefont {Johnson}, \citenamefont {Knopp}, \citenamefont {Lemke},
  \citenamefont {Milne}, \citenamefont {Pedrini}, \citenamefont {Radi},
  \citenamefont {Schertler}, \citenamefont {Standfuss}, \citenamefont {Staub},\
  and\ \citenamefont {Patthey}}]{SWISSFEL2017}%
  \BibitemOpen
  \bibfield  {author} {\bibinfo {author} {\bibfnamefont {R.}~\bibnamefont
  {Abela}}, \bibinfo {author} {\bibfnamefont {P.}~\bibnamefont {Beaud}},
  \bibinfo {author} {\bibfnamefont {J.~A.}\ \bibnamefont {van Bokhoven}},
  \bibinfo {author} {\bibfnamefont {M.}~\bibnamefont {Chergui}}, \bibinfo
  {author} {\bibfnamefont {T.}~\bibnamefont {Feurer}}, \bibinfo {author}
  {\bibfnamefont {J.}~\bibnamefont {Haase}}, \bibinfo {author} {\bibfnamefont
  {G.}~\bibnamefont {Ingold}}, \bibinfo {author} {\bibfnamefont {S.~L.}\
  \bibnamefont {Johnson}}, \bibinfo {author} {\bibfnamefont {G.}~\bibnamefont
  {Knopp}}, \bibinfo {author} {\bibfnamefont {H.}~\bibnamefont {Lemke}},
  \bibinfo {author} {\bibfnamefont {C.~J.}\ \bibnamefont {Milne}}, \bibinfo
  {author} {\bibfnamefont {B.}~\bibnamefont {Pedrini}}, \bibinfo {author}
  {\bibfnamefont {P.}~\bibnamefont {Radi}}, \bibinfo {author} {\bibfnamefont
  {G.}~\bibnamefont {Schertler}}, \bibinfo {author} {\bibfnamefont
  {J.}~\bibnamefont {Standfuss}}, \bibinfo {author} {\bibfnamefont
  {U.}~\bibnamefont {Staub}}, \ and\ \bibinfo {author} {\bibfnamefont
  {L.}~\bibnamefont {Patthey}},\ }\href@noop {} {\bibfield  {journal} {\bibinfo
   {journal} {Structural Dynamics}\ }\textbf {\bibinfo {volume} {4}},\ \bibinfo
  {pages} {061602}}\BibitemShut {NoStop}%
\bibitem [{\citenamefont {Lovesey}(1986)}]{LOVESEY}%
  \BibitemOpen
  \bibfield  {author} {\bibinfo {author} {\bibfnamefont {S.~W.}\ \bibnamefont
  {Lovesey}},\ }\href@noop {} {\emph {\bibinfo {title} {Theory of Neutron
  Scattering from Condensed Matter - Volume I: Nuclear Scattering}}}\ (\bibinfo
   {publisher} {Oxford University Press},\ \bibinfo {year} {1986})\BibitemShut
  {NoStop}%
\bibitem [{\citenamefont {Baron}\ \emph {et~al.}(1997)\citenamefont {Baron},
  \citenamefont {Franz}, \citenamefont {Meyer}, \citenamefont {R\"uffer},
  \citenamefont {Chumakov}, \citenamefont {Burkel},\ and\ \citenamefont
  {Petry}}]{BARON}%
  \BibitemOpen
  \bibfield  {author} {\bibinfo {author} {\bibfnamefont {A.~Q.~R.}\
  \bibnamefont {Baron}}, \bibinfo {author} {\bibfnamefont {H.}~\bibnamefont
  {Franz}}, \bibinfo {author} {\bibfnamefont {A.}~\bibnamefont {Meyer}},
  \bibinfo {author} {\bibfnamefont {R.}~\bibnamefont {R\"uffer}}, \bibinfo
  {author} {\bibfnamefont {A.~I.}\ \bibnamefont {Chumakov}}, \bibinfo {author}
  {\bibfnamefont {E.}~\bibnamefont {Burkel}}, \ and\ \bibinfo {author}
  {\bibfnamefont {W.}~\bibnamefont {Petry}},\ }\href@noop {} {\bibfield
  {journal} {\bibinfo  {journal} {Phys. Rev. Lett.}\ }\textbf {\bibinfo
  {volume} {79}},\ \bibinfo {pages} {2823} (\bibinfo {year}
  {1997})}\BibitemShut {NoStop}%
\bibitem [{\citenamefont {Smirnov}\ \emph {et~al.}(2001)\citenamefont
  {Smirnov}, \citenamefont {Kohn},\ and\ \citenamefont {Petry}}]{SMIRNOV}%
  \BibitemOpen
  \bibfield  {author} {\bibinfo {author} {\bibfnamefont {G.~V.}\ \bibnamefont
  {Smirnov}}, \bibinfo {author} {\bibfnamefont {V.~G.}\ \bibnamefont {Kohn}}, \
  and\ \bibinfo {author} {\bibfnamefont {W.}~\bibnamefont {Petry}},\
  }\href@noop {} {\bibfield  {journal} {\bibinfo  {journal} {Phys. Rev. B}\
  }\textbf {\bibinfo {volume} {63}},\ \bibinfo {pages} {144303} (\bibinfo
  {year} {2001})}\BibitemShut {NoStop}%
\bibitem [{\citenamefont {Smirnov}\ \emph {et~al.}(2006)\citenamefont
  {Smirnov}, \citenamefont {van B\"urck}, \citenamefont {Franz}, \citenamefont
  {Asthalter}, \citenamefont {Leupold}, \citenamefont {Schreier},\ and\
  \citenamefont {Petry}}]{SMIRNOV2006}%
  \BibitemOpen
  \bibfield  {author} {\bibinfo {author} {\bibfnamefont {G.~V.}\ \bibnamefont
  {Smirnov}}, \bibinfo {author} {\bibfnamefont {U.}~\bibnamefont {van
  B\"urck}}, \bibinfo {author} {\bibfnamefont {H.}~\bibnamefont {Franz}},
  \bibinfo {author} {\bibfnamefont {T.}~\bibnamefont {Asthalter}}, \bibinfo
  {author} {\bibfnamefont {O.}~\bibnamefont {Leupold}}, \bibinfo {author}
  {\bibfnamefont {E.}~\bibnamefont {Schreier}}, \ and\ \bibinfo {author}
  {\bibfnamefont {W.}~\bibnamefont {Petry}},\ }\href@noop {} {\bibfield
  {journal} {\bibinfo  {journal} {Phys. Rev. B}\ }\textbf {\bibinfo {volume}
  {73}},\ \bibinfo {pages} {184126} (\bibinfo {year} {2006})}\BibitemShut
  {NoStop}%
\bibitem [{\citenamefont {Saito}\ \emph
  {et~al.}(2012{\natexlab{a}})\citenamefont {Saito}, \citenamefont {Seto},
  \citenamefont {Kitao}, \citenamefont {Kobayashi}, \citenamefont {Kurokuzu},\
  and\ \citenamefont {Yoda}}]{Saito2012}%
  \BibitemOpen
  \bibfield  {author} {\bibinfo {author} {\bibfnamefont {M.}~\bibnamefont
  {Saito}}, \bibinfo {author} {\bibfnamefont {M.}~\bibnamefont {Seto}},
  \bibinfo {author} {\bibfnamefont {S.}~\bibnamefont {Kitao}}, \bibinfo
  {author} {\bibfnamefont {Y.}~\bibnamefont {Kobayashi}}, \bibinfo {author}
  {\bibfnamefont {M.}~\bibnamefont {Kurokuzu}}, \ and\ \bibinfo {author}
  {\bibfnamefont {Y.}~\bibnamefont {Yoda}},\ }\href@noop {} {\bibfield
  {journal} {\bibinfo  {journal} {Hyperfine Interactions}\ }\textbf {\bibinfo
  {volume} {206}},\ \bibinfo {pages} {87} (\bibinfo {year}
  {2012}{\natexlab{a}})}\BibitemShut {NoStop}%
\bibitem [{\citenamefont {Saito}\ \emph {et~al.}(2017)\citenamefont {Saito},
  \citenamefont {Masuda}, \citenamefont {Yoda},\ and\ \citenamefont
  {Seto}}]{Saito2017}%
  \BibitemOpen
  \bibfield  {author} {\bibinfo {author} {\bibfnamefont {M.}~\bibnamefont
  {Saito}}, \bibinfo {author} {\bibfnamefont {R.}~\bibnamefont {Masuda}},
  \bibinfo {author} {\bibfnamefont {Y.}~\bibnamefont {Yoda}}, \ and\ \bibinfo
  {author} {\bibfnamefont {M.}~\bibnamefont {Seto}},\ }\href@noop {} {\bibfield
   {journal} {\bibinfo  {journal} {Nature Sci. Rep.}\ }\textbf {\bibinfo
  {volume} {7}} (\bibinfo {year} {2017})}\BibitemShut {NoStop}%
\bibitem [{\citenamefont {Kaisermayr}\ \emph {et~al.}(2001)\citenamefont
  {Kaisermayr}, \citenamefont {Sepiol}, \citenamefont {Thiess}, \citenamefont
  {Vogl}, \citenamefont {Alp},\ and\ \citenamefont
  {Sturhahn}}]{Kaisermayr2001}%
  \BibitemOpen
  \bibfield  {author} {\bibinfo {author} {\bibfnamefont {M.}~\bibnamefont
  {Kaisermayr}}, \bibinfo {author} {\bibfnamefont {B.}~\bibnamefont {Sepiol}},
  \bibinfo {author} {\bibfnamefont {H.}~\bibnamefont {Thiess}}, \bibinfo
  {author} {\bibfnamefont {G.}~\bibnamefont {Vogl}}, \bibinfo {author}
  {\bibfnamefont {E.}~\bibnamefont {Alp}}, \ and\ \bibinfo {author}
  {\bibfnamefont {W.}~\bibnamefont {Sturhahn}},\ }\href@noop {} {\bibfield
  {journal} {\bibinfo  {journal} {Eur. Phys. J. B}\ }\textbf {\bibinfo {volume}
  {20}},\ \bibinfo {pages} {335} (\bibinfo {year} {2001})}\BibitemShut
  {NoStop}%
\bibitem [{\citenamefont {Saito}\ \emph {et~al.}(2009)\citenamefont {Saito},
  \citenamefont {Seto}, \citenamefont {Kitao}, \citenamefont {Kobayashi},
  \citenamefont {Higashitaniguchi}, \citenamefont {Kurokuzu}, \citenamefont
  {Sugiyama},\ and\ \citenamefont {Yoda}}]{1882-0786-2-2-026502}%
  \BibitemOpen
  \bibfield  {author} {\bibinfo {author} {\bibfnamefont {M.}~\bibnamefont
  {Saito}}, \bibinfo {author} {\bibfnamefont {M.}~\bibnamefont {Seto}},
  \bibinfo {author} {\bibfnamefont {S.}~\bibnamefont {Kitao}}, \bibinfo
  {author} {\bibfnamefont {Y.}~\bibnamefont {Kobayashi}}, \bibinfo {author}
  {\bibfnamefont {S.}~\bibnamefont {Higashitaniguchi}}, \bibinfo {author}
  {\bibfnamefont {M.}~\bibnamefont {Kurokuzu}}, \bibinfo {author}
  {\bibfnamefont {M.}~\bibnamefont {Sugiyama}}, \ and\ \bibinfo {author}
  {\bibfnamefont {Y.}~\bibnamefont {Yoda}},\ }\href@noop {} {\bibfield
  {journal} {\bibinfo  {journal} {Appl. Phys. Express}\ }\textbf {\bibinfo
  {volume} {2}},\ \bibinfo {pages} {026502} (\bibinfo {year}
  {2009})}\BibitemShut {NoStop}%
\bibitem [{\citenamefont {Chumakov}\ \emph {et~al.}(2018)\citenamefont
  {Chumakov}, \citenamefont {Baron}, \citenamefont {Sergueev}, \citenamefont
  {Strohm}, \citenamefont {Leupold}, \citenamefont {Shvyd’ko}, \citenamefont
  {Smirnov}, \citenamefont {R\"uffer}, \citenamefont {Inubushi}, \citenamefont
  {Yabashi}, \citenamefont {Tono}, \citenamefont {Kudo},\ and\ \citenamefont
  {Ishikawa}}]{chumakov}%
  \BibitemOpen
  \bibfield  {author} {\bibinfo {author} {\bibfnamefont {A.~I.}\ \bibnamefont
  {Chumakov}}, \bibinfo {author} {\bibfnamefont {A.~Q.~R.}\ \bibnamefont
  {Baron}}, \bibinfo {author} {\bibfnamefont {I.}~\bibnamefont {Sergueev}},
  \bibinfo {author} {\bibfnamefont {C.}~\bibnamefont {Strohm}}, \bibinfo
  {author} {\bibfnamefont {O.}~\bibnamefont {Leupold}}, \bibinfo {author}
  {\bibfnamefont {Y.}~\bibnamefont {Shvyd’ko}}, \bibinfo {author}
  {\bibfnamefont {G.~V.}\ \bibnamefont {Smirnov}}, \bibinfo {author}
  {\bibfnamefont {R.}~\bibnamefont {R\"uffer}}, \bibinfo {author}
  {\bibfnamefont {Y.}~\bibnamefont {Inubushi}}, \bibinfo {author}
  {\bibfnamefont {M.}~\bibnamefont {Yabashi}}, \bibinfo {author} {\bibfnamefont
  {K.}~\bibnamefont {Tono}}, \bibinfo {author} {\bibfnamefont {T.}~\bibnamefont
  {Kudo}}, \ and\ \bibinfo {author} {\bibfnamefont {T.}~\bibnamefont
  {Ishikawa}},\ }\href@noop {} {\bibfield  {journal} {\bibinfo  {journal}
  {Nature Phys.}\ }\textbf {\bibinfo {volume} {14}},\ \bibinfo {pages} {261}
  (\bibinfo {year} {2018})}\BibitemShut {NoStop}%
\bibitem [{\citenamefont {Caporaletti}\ \emph {et~al.}(2017)\citenamefont
  {Caporaletti}, \citenamefont {Chumakov}, \citenamefont {R\"uffer},\ and\
  \citenamefont {Monaco}}]{doi:10.1063/1.5008868}%
  \BibitemOpen
  \bibfield  {author} {\bibinfo {author} {\bibfnamefont {F.}~\bibnamefont
  {Caporaletti}}, \bibinfo {author} {\bibfnamefont {A.~I.}\ \bibnamefont
  {Chumakov}}, \bibinfo {author} {\bibfnamefont {R.}~\bibnamefont {R\"uffer}},
  \ and\ \bibinfo {author} {\bibfnamefont {G.}~\bibnamefont {Monaco}},\
  }\href@noop {} {\bibfield  {journal} {\bibinfo  {journal} {Review of
  Scientific Instruments}\ }\textbf {\bibinfo {volume} {88}},\ \bibinfo {pages}
  {105114} (\bibinfo {year} {2017})}\BibitemShut {NoStop}%
\bibitem [{\citenamefont {Heeg}\ \emph {et~al.}(2017)\citenamefont {Heeg},
  \citenamefont {Kaldun}, \citenamefont {Strohm}, \citenamefont {Reiser},
  \citenamefont {Ott}, \citenamefont {Subramanian}, \citenamefont {Lentrodt},
  \citenamefont {Haber}, \citenamefont {Wille}, \citenamefont {Goerttler},
  \citenamefont {R{\"u}ffer}, \citenamefont {Keitel}, \citenamefont
  {R{\"o}hlsberger}, \citenamefont {Pfeifer},\ and\ \citenamefont
  {Evers}}]{Heeg2017}%
  \BibitemOpen
  \bibfield  {author} {\bibinfo {author} {\bibfnamefont {K.~P.}\ \bibnamefont
  {Heeg}}, \bibinfo {author} {\bibfnamefont {A.}~\bibnamefont {Kaldun}},
  \bibinfo {author} {\bibfnamefont {C.}~\bibnamefont {Strohm}}, \bibinfo
  {author} {\bibfnamefont {P.}~\bibnamefont {Reiser}}, \bibinfo {author}
  {\bibfnamefont {C.}~\bibnamefont {Ott}}, \bibinfo {author} {\bibfnamefont
  {R.}~\bibnamefont {Subramanian}}, \bibinfo {author} {\bibfnamefont
  {D.}~\bibnamefont {Lentrodt}}, \bibinfo {author} {\bibfnamefont
  {J.}~\bibnamefont {Haber}}, \bibinfo {author} {\bibfnamefont {H.-C.}\
  \bibnamefont {Wille}}, \bibinfo {author} {\bibfnamefont {S.}~\bibnamefont
  {Goerttler}}, \bibinfo {author} {\bibfnamefont {R.}~\bibnamefont
  {R{\"u}ffer}}, \bibinfo {author} {\bibfnamefont {C.~H.}\ \bibnamefont
  {Keitel}}, \bibinfo {author} {\bibfnamefont {R.}~\bibnamefont
  {R{\"o}hlsberger}}, \bibinfo {author} {\bibfnamefont {T.}~\bibnamefont
  {Pfeifer}}, \ and\ \bibinfo {author} {\bibfnamefont {J.}~\bibnamefont
  {Evers}},\ }\href {\doibase 10.1126/science.aan3512} {\bibfield  {journal}
  {\bibinfo  {journal} {Science}\ }\textbf {\bibinfo {volume} {357}},\ \bibinfo
  {pages} {375} (\bibinfo {year} {2017})}\BibitemShut {NoStop}%
\bibitem [{\citenamefont {Saito}\ \emph
  {et~al.}(2012{\natexlab{b}})\citenamefont {Saito}, \citenamefont {Kitao},
  \citenamefont {Kobayashi}, \citenamefont {Kurokuzu}, \citenamefont {Yoda},\
  and\ \citenamefont {Seto}}]{Saito2012bis}%
  \BibitemOpen
  \bibfield  {author} {\bibinfo {author} {\bibfnamefont {M.}~\bibnamefont
  {Saito}}, \bibinfo {author} {\bibfnamefont {S.}~\bibnamefont {Kitao}},
  \bibinfo {author} {\bibfnamefont {Y.}~\bibnamefont {Kobayashi}}, \bibinfo
  {author} {\bibfnamefont {M.}~\bibnamefont {Kurokuzu}}, \bibinfo {author}
  {\bibfnamefont {Y.}~\bibnamefont {Yoda}}, \ and\ \bibinfo {author}
  {\bibfnamefont {M.}~\bibnamefont {Seto}},\ }\href@noop {} {\bibfield
  {journal} {\bibinfo  {journal} {Phys. Rev. Lett.}\ }\textbf {\bibinfo
  {volume} {109}},\ \bibinfo {pages} {115705} (\bibinfo {year}
  {2012}{\natexlab{b}})}\BibitemShut {NoStop}%
\bibitem [{\citenamefont {Saito}\ \emph {et~al.}(2014)\citenamefont {Saito},
  \citenamefont {Battistoni}, \citenamefont {Kitao}, \citenamefont {Kobayashi},
  \citenamefont {Kurokuzu}, \citenamefont {Yoda},\ and\ \citenamefont
  {Seto}}]{Saito2014}%
  \BibitemOpen
  \bibfield  {author} {\bibinfo {author} {\bibfnamefont {M.}~\bibnamefont
  {Saito}}, \bibinfo {author} {\bibfnamefont {A.}~\bibnamefont {Battistoni}},
  \bibinfo {author} {\bibfnamefont {S.}~\bibnamefont {Kitao}}, \bibinfo
  {author} {\bibfnamefont {Y.}~\bibnamefont {Kobayashi}}, \bibinfo {author}
  {\bibfnamefont {M.}~\bibnamefont {Kurokuzu}}, \bibinfo {author}
  {\bibfnamefont {Y.}~\bibnamefont {Yoda}}, \ and\ \bibinfo {author}
  {\bibfnamefont {M.}~\bibnamefont {Seto}},\ }\href@noop {} {\bibfield
  {journal} {\bibinfo  {journal} {Hyperfine Interactions}\ }\textbf {\bibinfo
  {volume} {226}},\ \bibinfo {pages} {629} (\bibinfo {year}
  {2014})}\BibitemShut {NoStop}%
\bibitem [{\citenamefont {Hansen}\ and\ \citenamefont
  {McDonald}(2013)}]{HANSEN}%
  \BibitemOpen
  \bibfield  {author} {\bibinfo {author} {\bibfnamefont {J.~P.}\ \bibnamefont
  {Hansen}}\ and\ \bibinfo {author} {\bibfnamefont {I.~R.}\ \bibnamefont
  {McDonald}},\ }\href@noop {} {\emph {\bibinfo {title} {Theory of Simple
  Liquids}}}\ (\bibinfo  {publisher} {Academic Press},\ \bibinfo {year}
  {2013})\BibitemShut {NoStop}%
\bibitem [{\citenamefont {Van~Hove}(1958)}]{VANHOVE1}%
  \BibitemOpen
  \bibfield  {author} {\bibinfo {author} {\bibfnamefont {L.}~\bibnamefont
  {Van~Hove}},\ }\href@noop {} {\bibfield  {journal} {\bibinfo  {journal}
  {Physica XXIV Zernike issue}\ }\textbf {\bibinfo {volume} {24}},\ \bibinfo
  {pages} {404} (\bibinfo {year} {1958})}\BibitemShut {NoStop}%
\bibitem [{\citenamefont {Watson}(1996)}]{Watson}%
  \BibitemOpen
  \bibfield  {author} {\bibinfo {author} {\bibfnamefont {G.~I.}\ \bibnamefont
  {Watson}},\ }\href {http://stacks.iop.org/0953-8984/8/i=33/a=005} {\bibfield
  {journal} {\bibinfo  {journal} {Journal of Physics: Condensed Matter}\
  }\textbf {\bibinfo {volume} {8}},\ \bibinfo {pages} {5955} (\bibinfo {year}
  {1996})}\BibitemShut {NoStop}%
\bibitem [{\citenamefont {Barocchi}\ \emph {et~al.}(1982)\citenamefont
  {Barocchi}, \citenamefont {Moraldi},\ and\ \citenamefont {Zoppi}}]{barocchi}%
  \BibitemOpen
  \bibfield  {author} {\bibinfo {author} {\bibfnamefont {F.}~\bibnamefont
  {Barocchi}}, \bibinfo {author} {\bibfnamefont {M.}~\bibnamefont {Moraldi}}, \
  and\ \bibinfo {author} {\bibfnamefont {M.}~\bibnamefont {Zoppi}},\
  }\href@noop {} {\bibfield  {journal} {\bibinfo  {journal} {Phys. Rev. A}\
  }\textbf {\bibinfo {volume} {26}},\ \bibinfo {pages} {2168} (\bibinfo {year}
  {1982})}\BibitemShut {NoStop}%
\bibitem [{\citenamefont {Sears}(1985)}]{sears}%
  \BibitemOpen
  \bibfield  {author} {\bibinfo {author} {\bibfnamefont {V.~F.}\ \bibnamefont
  {Sears}},\ }\href@noop {} {\bibfield  {journal} {\bibinfo  {journal} {Phys.
  Rev. A}\ }\textbf {\bibinfo {volume} {31}},\ \bibinfo {pages} {2525}
  (\bibinfo {year} {1985})}\BibitemShut {NoStop}%
\bibitem [{\citenamefont {Schofield}(1960)}]{schofield}%
  \BibitemOpen
  \bibfield  {author} {\bibinfo {author} {\bibfnamefont {P.}~\bibnamefont
  {Schofield}},\ }\href {\doibase 10.1103/PhysRevLett.4.239} {\bibfield
  {journal} {\bibinfo  {journal} {Phys. Rev. Lett.}\ }\textbf {\bibinfo
  {volume} {4}},\ \bibinfo {pages} {239} (\bibinfo {year} {1960})}\BibitemShut
  {NoStop}%
\bibitem [{\citenamefont {Cunsolo}\ \emph {et~al.}(2005)\citenamefont
  {Cunsolo}, \citenamefont {Colognesi}, \citenamefont {Sampoli}, \citenamefont
  {Senesi},\ and\ \citenamefont {Verbeni}}]{cunsolo05}%
  \BibitemOpen
  \bibfield  {author} {\bibinfo {author} {\bibfnamefont {A.}~\bibnamefont
  {Cunsolo}}, \bibinfo {author} {\bibfnamefont {D.}~\bibnamefont {Colognesi}},
  \bibinfo {author} {\bibfnamefont {M.}~\bibnamefont {Sampoli}}, \bibinfo
  {author} {\bibfnamefont {R.}~\bibnamefont {Senesi}}, \ and\ \bibinfo {author}
  {\bibfnamefont {R.}~\bibnamefont {Verbeni}},\ }\href@noop {} {\bibfield
  {journal} {\bibinfo  {journal} {The Journal of Chemical Physics}\ }\textbf
  {\bibinfo {volume} {123}},\ \bibinfo {pages} {114509} (\bibinfo {year}
  {2005})}\BibitemShut {NoStop}%
\bibitem [{\citenamefont {Cunsolo}\ \emph {et~al.}(2003)\citenamefont
  {Cunsolo}, \citenamefont {Monaco}, \citenamefont {Nardone}, \citenamefont
  {Pratesi},\ and\ \citenamefont {Verbeni}}]{cunsolo03}%
  \BibitemOpen
  \bibfield  {author} {\bibinfo {author} {\bibfnamefont {A.}~\bibnamefont
  {Cunsolo}}, \bibinfo {author} {\bibfnamefont {G.}~\bibnamefont {Monaco}},
  \bibinfo {author} {\bibfnamefont {M.}~\bibnamefont {Nardone}}, \bibinfo
  {author} {\bibfnamefont {G.}~\bibnamefont {Pratesi}}, \ and\ \bibinfo
  {author} {\bibfnamefont {R.}~\bibnamefont {Verbeni}},\ }\href {\doibase
  10.1103/PhysRevB.67.024507} {\bibfield  {journal} {\bibinfo  {journal} {Phys.
  Rev. B}\ }\textbf {\bibinfo {volume} {67}},\ \bibinfo {pages} {024507}
  (\bibinfo {year} {2003})}\BibitemShut {NoStop}%
\bibitem [{\citenamefont {Jardine}\ \emph {et~al.}(2009)\citenamefont
  {Jardine}, \citenamefont {Alexandrowicz}, \citenamefont {Hedgeland},
  \citenamefont {Allison},\ and\ \citenamefont {Ellis}}]{surface1}%
  \BibitemOpen
  \bibfield  {author} {\bibinfo {author} {\bibfnamefont {A.~P.}\ \bibnamefont
  {Jardine}}, \bibinfo {author} {\bibfnamefont {G.}~\bibnamefont
  {Alexandrowicz}}, \bibinfo {author} {\bibfnamefont {H.}~\bibnamefont
  {Hedgeland}}, \bibinfo {author} {\bibfnamefont {W.}~\bibnamefont {Allison}},
  \ and\ \bibinfo {author} {\bibfnamefont {J.}~\bibnamefont {Ellis}},\ }\href
  {\doibase 10.1039/B810769F} {\bibfield  {journal} {\bibinfo  {journal} {Phys.
  Chem. Chem. Phys.}\ }\textbf {\bibinfo {volume} {11}},\ \bibinfo {pages}
  {3355} (\bibinfo {year} {2009})}\BibitemShut {NoStop}%
\bibitem [{\citenamefont {Gross}\ and\ \citenamefont {Bloch}(2017)}]{Gross995}%
  \BibitemOpen
  \bibfield  {author} {\bibinfo {author} {\bibfnamefont {C.}~\bibnamefont
  {Gross}}\ and\ \bibinfo {author} {\bibfnamefont {I.}~\bibnamefont {Bloch}},\
  }\href@noop {} {\bibfield  {journal} {\bibinfo  {journal} {Science}\ }\textbf
  {\bibinfo {volume} {357}},\ \bibinfo {pages} {995} (\bibinfo {year}
  {2017})}\BibitemShut {NoStop}%
\bibitem [{\citenamefont {Georgescu}\ \emph {et~al.}(2014)\citenamefont
  {Georgescu}, \citenamefont {Ashhab},\ and\ \citenamefont
  {Nori}}]{RevModPhys.86.153}%
  \BibitemOpen
  \bibfield  {author} {\bibinfo {author} {\bibfnamefont {I.~M.}\ \bibnamefont
  {Georgescu}}, \bibinfo {author} {\bibfnamefont {S.}~\bibnamefont {Ashhab}}, \
  and\ \bibinfo {author} {\bibfnamefont {F.}~\bibnamefont {Nori}},\ }\href
  {\doibase 10.1103/RevModPhys.86.153} {\bibfield  {journal} {\bibinfo
  {journal} {Rev. Mod. Phys.}\ }\textbf {\bibinfo {volume} {86}},\ \bibinfo
  {pages} {153} (\bibinfo {year} {2014})}\BibitemShut {NoStop}%
\bibitem [{\citenamefont {Landig}\ \emph {et~al.}(2015)\citenamefont {Landig},
  \citenamefont {Brennecke}, \citenamefont {Mottl}, \citenamefont {Donner},\
  and\ \citenamefont {Esslinger}}]{landig}%
  \BibitemOpen
  \bibfield  {author} {\bibinfo {author} {\bibfnamefont {R.}~\bibnamefont
  {Landig}}, \bibinfo {author} {\bibfnamefont {F.}~\bibnamefont {Brennecke}},
  \bibinfo {author} {\bibfnamefont {R.}~\bibnamefont {Mottl}}, \bibinfo
  {author} {\bibfnamefont {T.}~\bibnamefont {Donner}}, \ and\ \bibinfo {author}
  {\bibfnamefont {T.}~\bibnamefont {Esslinger}},\ }\href@noop {} {\bibfield
  {journal} {\bibinfo  {journal} {Nature Communications}\ }\textbf {\bibinfo
  {volume} {6}},\ \bibinfo {pages} {7046} (\bibinfo {year} {2015})}\BibitemShut
  {NoStop}%
\bibitem [{\citenamefont {Keimer}\ and\ \citenamefont {Moore}(2017)}]{keimer}%
  \BibitemOpen
  \bibfield  {author} {\bibinfo {author} {\bibfnamefont {B.}~\bibnamefont
  {Keimer}}\ and\ \bibinfo {author} {\bibfnamefont {J.~E.}\ \bibnamefont
  {Moore}},\ }\href@noop {} {\bibfield  {journal} {\bibinfo  {journal} {Nature
  Physics}\ }\textbf {\bibinfo {volume} {13}},\ \bibinfo {pages} {1045}
  (\bibinfo {year} {2017})}\BibitemShut {NoStop}%
\bibitem [{\citenamefont {Dagotto}(2005)}]{Dagotto257}%
  \BibitemOpen
  \bibfield  {author} {\bibinfo {author} {\bibfnamefont {E.}~\bibnamefont
  {Dagotto}},\ }\href {\doibase 10.1126/science.1107559} {\bibfield  {journal}
  {\bibinfo  {journal} {Science}\ }\textbf {\bibinfo {volume} {309}},\ \bibinfo
  {pages} {257} (\bibinfo {year} {2005})}\BibitemShut {NoStop}%
\bibitem [{\citenamefont {Quintanilla}\ and\ \citenamefont
  {Hooley}(2009)}]{QUINTANILLA}%
  \BibitemOpen
  \bibfield  {author} {\bibinfo {author} {\bibfnamefont {J.}~\bibnamefont
  {Quintanilla}}\ and\ \bibinfo {author} {\bibfnamefont {C.}~\bibnamefont
  {Hooley}},\ }\href@noop {} {\bibfield  {journal} {\bibinfo  {journal}
  {Physics World}\ }\textbf {\bibinfo {volume} {22}},\ \bibinfo {pages} {32}
  (\bibinfo {year} {2009})}\BibitemShut {NoStop}%
\bibitem [{\citenamefont {Schroer}\ \emph {et~al.}(2018)\citenamefont
  {Schroer}, \citenamefont {Agapov}, \citenamefont {Brefeld}, \citenamefont
  {Brinkmann}, \citenamefont {Chae}, \citenamefont {Chao}, \citenamefont
  {Eriksson}, \citenamefont {Keil}, \citenamefont {N.~Gavald{\`{a}}},
  \citenamefont {R{\"{o}}hlsberger}, \citenamefont {Seeck}, \citenamefont
  {Sprung}, \citenamefont {Tischer}, \citenamefont {Wanzenberg},\ and\
  \citenamefont {Weckert}}]{petra4}%
  \BibitemOpen
  \bibfield  {author} {\bibinfo {author} {\bibfnamefont {C.~G.}\ \bibnamefont
  {Schroer}}, \bibinfo {author} {\bibfnamefont {I.}~\bibnamefont {Agapov}},
  \bibinfo {author} {\bibfnamefont {W.}~\bibnamefont {Brefeld}}, \bibinfo
  {author} {\bibfnamefont {R.}~\bibnamefont {Brinkmann}}, \bibinfo {author}
  {\bibfnamefont {Y.-C.}\ \bibnamefont {Chae}}, \bibinfo {author}
  {\bibfnamefont {H.-C.}\ \bibnamefont {Chao}}, \bibinfo {author}
  {\bibfnamefont {M.}~\bibnamefont {Eriksson}}, \bibinfo {author}
  {\bibfnamefont {J.}~\bibnamefont {Keil}}, \bibinfo {author} {\bibfnamefont
  {X.}~\bibnamefont {N.~Gavald{\`{a}}}}, \bibinfo {author} {\bibfnamefont
  {R.}~\bibnamefont {R{\"{o}}hlsberger}}, \bibinfo {author} {\bibfnamefont
  {O.~H.}\ \bibnamefont {Seeck}}, \bibinfo {author} {\bibfnamefont
  {M.}~\bibnamefont {Sprung}}, \bibinfo {author} {\bibfnamefont
  {M.}~\bibnamefont {Tischer}}, \bibinfo {author} {\bibfnamefont
  {R.}~\bibnamefont {Wanzenberg}}, \ and\ \bibinfo {author} {\bibfnamefont
  {E.}~\bibnamefont {Weckert}},\ }\href@noop {} {\bibfield  {journal} {\bibinfo
   {journal} {Journal of Synchrotron Radiation}\ }\textbf {\bibinfo {volume}
  {25}},\ \bibinfo {pages} {1277} (\bibinfo {year} {2018})}\BibitemShut
  {NoStop}%
\bibitem [{\citenamefont {Seaberg}\ \emph {et~al.}(2017)\citenamefont
  {Seaberg}, \citenamefont {Holladay}, \citenamefont {Lee}, \citenamefont
  {Sikorski}, \citenamefont {Reid}, \citenamefont {Montoya}, \citenamefont
  {Dakovski}, \citenamefont {Koralek}, \citenamefont {Coslovich}, \citenamefont
  {Moeller}, \citenamefont {Schlotter}, \citenamefont {Streubel}, \citenamefont
  {Kevan}, \citenamefont {Fischer}, \citenamefont {Fullerton}, \citenamefont
  {Turner}, \citenamefont {Decker}, \citenamefont {Sinha}, \citenamefont
  {Roy},\ and\ \citenamefont {Turner}}]{xpcs}%
  \BibitemOpen
  \bibfield  {author} {\bibinfo {author} {\bibfnamefont {M.~H.}\ \bibnamefont
  {Seaberg}}, \bibinfo {author} {\bibfnamefont {B.}~\bibnamefont {Holladay}},
  \bibinfo {author} {\bibfnamefont {J.~C.~T.}\ \bibnamefont {Lee}}, \bibinfo
  {author} {\bibfnamefont {M.}~\bibnamefont {Sikorski}}, \bibinfo {author}
  {\bibfnamefont {A.~H.}\ \bibnamefont {Reid}}, \bibinfo {author}
  {\bibfnamefont {S.~A.}\ \bibnamefont {Montoya}}, \bibinfo {author}
  {\bibfnamefont {G.~L.}\ \bibnamefont {Dakovski}}, \bibinfo {author}
  {\bibfnamefont {J.~D.}\ \bibnamefont {Koralek}}, \bibinfo {author}
  {\bibfnamefont {G.}~\bibnamefont {Coslovich}}, \bibinfo {author}
  {\bibfnamefont {S.}~\bibnamefont {Moeller}}, \bibinfo {author} {\bibfnamefont
  {W.~F.}\ \bibnamefont {Schlotter}}, \bibinfo {author} {\bibfnamefont
  {R.}~\bibnamefont {Streubel}}, \bibinfo {author} {\bibfnamefont {S.~D.}\
  \bibnamefont {Kevan}}, \bibinfo {author} {\bibfnamefont {P.}~\bibnamefont
  {Fischer}}, \bibinfo {author} {\bibfnamefont {E.~E.}\ \bibnamefont
  {Fullerton}}, \bibinfo {author} {\bibfnamefont {J.~L.}\ \bibnamefont
  {Turner}}, \bibinfo {author} {\bibfnamefont {F.-J.}\ \bibnamefont {Decker}},
  \bibinfo {author} {\bibfnamefont {S.~K.}\ \bibnamefont {Sinha}}, \bibinfo
  {author} {\bibfnamefont {S.}~\bibnamefont {Roy}}, \ and\ \bibinfo {author}
  {\bibfnamefont {J.~J.}\ \bibnamefont {Turner}},\ }\href@noop {} {\bibfield
  {journal} {\bibinfo  {journal} {Phys. Rev. Lett.}\ }\textbf {\bibinfo
  {volume} {119}},\ \bibinfo {pages} {067403} (\bibinfo {year}
  {2017})}\BibitemShut {NoStop}%
\bibitem [{\citenamefont {Vagizov}\ \emph {et~al.}(2014)\citenamefont
  {Vagizov}, \citenamefont {Antonov}, \citenamefont {Radeonychev},
  \citenamefont {Shakhmuratov},\ and\ \citenamefont
  {Kocharovskaya}}]{Vagizov2014}%
  \BibitemOpen
  \bibfield  {author} {\bibinfo {author} {\bibfnamefont {F.}~\bibnamefont
  {Vagizov}}, \bibinfo {author} {\bibfnamefont {V.}~\bibnamefont {Antonov}},
  \bibinfo {author} {\bibfnamefont {Y.~V.}\ \bibnamefont {Radeonychev}},
  \bibinfo {author} {\bibfnamefont {R.~N.}\ \bibnamefont {Shakhmuratov}}, \
  and\ \bibinfo {author} {\bibfnamefont {O.}~\bibnamefont {Kocharovskaya}},\
  }\href {http://dx.doi.org/10.1038/nature13018} {\bibfield  {journal}
  {\bibinfo  {journal} {Nature}\ }\textbf {\bibinfo {volume} {508}},\ \bibinfo
  {pages} {80} (\bibinfo {year} {2014})}\BibitemShut {NoStop}%
\bibitem [{sup()}]{supplement}%
  \BibitemOpen
  \href@noop {} {\enquote {\bibinfo {title} {Supplemental material to ``probing
  quantum dynamical couple correlations with time-domain interferometry"},}\
  }\BibitemShut {NoStop}%
\bibitem [{\citenamefont {Sakshath}\ \emph {et~al.}(2017)\citenamefont
  {Sakshath}, \citenamefont {Jenni}, \citenamefont {Scherthan}, \citenamefont
  {W{\"u}rtz}, \citenamefont {Herlitschke}, \citenamefont {Sergeev},
  \citenamefont {Strohm}, \citenamefont {Wille}, \citenamefont
  {R{\"o}hlsberger}, \citenamefont {Wolny},\ and\ \citenamefont
  {Sch{\"u}nemann}}]{Sakshath2017}%
  \BibitemOpen
  \bibfield  {author} {\bibinfo {author} {\bibfnamefont {S.}~\bibnamefont
  {Sakshath}}, \bibinfo {author} {\bibfnamefont {K.}~\bibnamefont {Jenni}},
  \bibinfo {author} {\bibfnamefont {L.}~\bibnamefont {Scherthan}}, \bibinfo
  {author} {\bibfnamefont {P.}~\bibnamefont {W{\"u}rtz}}, \bibinfo {author}
  {\bibfnamefont {M.}~\bibnamefont {Herlitschke}}, \bibinfo {author}
  {\bibfnamefont {I.}~\bibnamefont {Sergeev}}, \bibinfo {author} {\bibfnamefont
  {C.}~\bibnamefont {Strohm}}, \bibinfo {author} {\bibfnamefont {H.-C.}\
  \bibnamefont {Wille}}, \bibinfo {author} {\bibfnamefont {R.}~\bibnamefont
  {R{\"o}hlsberger}}, \bibinfo {author} {\bibfnamefont {J.~A.}\ \bibnamefont
  {Wolny}}, \ and\ \bibinfo {author} {\bibfnamefont {V.}~\bibnamefont
  {Sch{\"u}nemann}},\ }\href {\doibase 10.1007/s10751-017-1461-3} {\bibfield
  {journal} {\bibinfo  {journal} {Hyperfine Interactions}\ }\textbf {\bibinfo
  {volume} {238}},\ \bibinfo {pages} {89} (\bibinfo {year} {2017})}\BibitemShut
  {NoStop}%
\bibitem [{\citenamefont {Murmann}\ \emph {et~al.}(2015)\citenamefont
  {Murmann}, \citenamefont {Bergschneider}, \citenamefont {Klinkhamer},
  \citenamefont {Z\"urn}, \citenamefont {Lompe},\ and\ \citenamefont
  {Jochim}}]{MURMANN}%
  \BibitemOpen
  \bibfield  {author} {\bibinfo {author} {\bibfnamefont {S.}~\bibnamefont
  {Murmann}}, \bibinfo {author} {\bibfnamefont {A.}~\bibnamefont
  {Bergschneider}}, \bibinfo {author} {\bibfnamefont {V.~M.}\ \bibnamefont
  {Klinkhamer}}, \bibinfo {author} {\bibfnamefont {G.}~\bibnamefont {Z\"urn}},
  \bibinfo {author} {\bibfnamefont {T.}~\bibnamefont {Lompe}}, \ and\ \bibinfo
  {author} {\bibfnamefont {S.}~\bibnamefont {Jochim}},\ }\href@noop {}
  {\bibfield  {journal} {\bibinfo  {journal} {Phys. Rev. Lett.}\ }\textbf
  {\bibinfo {volume} {114}},\ \bibinfo {pages} {080402} (\bibinfo {year}
  {2015})}\BibitemShut {NoStop}%
\bibitem [{\citenamefont {Ballentine}(1995)}]{BALLENTINE}%
  \BibitemOpen
  \bibfield  {author} {\bibinfo {author} {\bibfnamefont {L.~E.}\ \bibnamefont
  {Ballentine}},\ }\enquote {\bibinfo {title} {Fundamental problems in quantum
  physics},}\ \ (\bibinfo  {publisher} {Springer},\ \bibinfo {year} {1995})\
  pp.\ \bibinfo {pages} {15--28}\BibitemShut {NoStop}%
\bibitem [{\citenamefont {Margenau}\ and\ \citenamefont
  {Hill}(1961)}]{MARGENAU}%
  \BibitemOpen
  \bibfield  {author} {\bibinfo {author} {\bibfnamefont {H.}~\bibnamefont
  {Margenau}}\ and\ \bibinfo {author} {\bibfnamefont {R.~N.}\ \bibnamefont
  {Hill}},\ }\href@noop {} {\bibfield  {journal} {\bibinfo  {journal} {Progress
  of Theoretical Physics}\ }\textbf {\bibinfo {volume} {26}},\ \bibinfo {pages}
  {722} (\bibinfo {year} {1961})}\BibitemShut {NoStop}%
\bibitem [{\citenamefont {Uhrich}\ \emph {et~al.}(2017)\citenamefont {Uhrich},
  \citenamefont {Castrignano}, \citenamefont {Uys},\ and\ \citenamefont
  {Kastner}}]{PhysRevA.96.022127}%
  \BibitemOpen
  \bibfield  {author} {\bibinfo {author} {\bibfnamefont {P.}~\bibnamefont
  {Uhrich}}, \bibinfo {author} {\bibfnamefont {S.}~\bibnamefont {Castrignano}},
  \bibinfo {author} {\bibfnamefont {H.}~\bibnamefont {Uys}}, \ and\ \bibinfo
  {author} {\bibfnamefont {M.}~\bibnamefont {Kastner}},\ }\href@noop {}
  {\bibfield  {journal} {\bibinfo  {journal} {Phys. Rev. A}\ }\textbf {\bibinfo
  {volume} {96}},\ \bibinfo {pages} {022127} (\bibinfo {year}
  {2017})}\BibitemShut {NoStop}%
\bibitem [{\citenamefont {Hau-Riege}(2015)}]{HAURIEGE}%
  \BibitemOpen
  \bibfield  {author} {\bibinfo {author} {\bibfnamefont {S.~P.}\ \bibnamefont
  {Hau-Riege}},\ }\href@noop {} {\emph {\bibinfo {title} {Nonrelativistic
  Quantum X-Ray Physics}}}\ (\bibinfo  {publisher} {Wiley-VCH},\ \bibinfo
  {year} {2015})\BibitemShut {NoStop}%
\bibitem [{\citenamefont {Messiah}(1964)}]{MESSIAH}%
  \BibitemOpen
  \bibfield  {author} {\bibinfo {author} {\bibfnamefont {A.}~\bibnamefont
  {Messiah}},\ }\href@noop {} {\emph {\bibinfo {title} {Quantum Mechanics}}}\
  (\bibinfo  {publisher} {North-Holland Publishing Company},\ \bibinfo {year}
  {1964})\BibitemShut {NoStop}%
\bibitem [{\citenamefont {Scully}\ and\ \citenamefont
  {Zubairy}(1997)}]{SCULLY}%
  \BibitemOpen
  \bibfield  {author} {\bibinfo {author} {\bibfnamefont {M.~O.}\ \bibnamefont
  {Scully}}\ and\ \bibinfo {author} {\bibfnamefont {M.~S.}\ \bibnamefont
  {Zubairy}},\ }\href@noop {} {\emph {\bibinfo {title} {Quantum Optics}}}\
  (\bibinfo  {publisher} {Cambridge University Press},\ \bibinfo {year}
  {1997})\BibitemShut {NoStop}%
\end{thebibliography}%

\clearpage
\onecolumngrid
\appendix

\begin{center}
{ \bf \large Supplemental Material to ``Probing Quantum Dynamical Couple Correlations\\with Time-Domain Interferometry"}
\end{center}

\section{Derivation of the single photon detection amplitudes}
In order to derive the explicit form for the detection amplitude, we need to solve the problem of the scattering of the photon by the target system after it passed through the split stage. Next, we have to evaluate how the initial state of the system composed of target and photon evolves for long times under the influence of the matter-radiation interaction. At the initial time after the split unit, the photon is in a state given by the superposition of two spatially separated incoming wave packets, $\ket{\gamma_1}+\ket{\gamma_2}$, whereas the target system is in the state $\ket{\psi}$, so that the initial state of the global system is the product of the two
\begin{equation}
\label{eqn: SMInitState}
\ket{\Psi}=\ket{\gamma_1}\ket{\psi}+\ket{\gamma_2}\ket{\psi}\,.
\end{equation}
Let us fix a reference frame with the $x$-axis parallel to the initial direction of propagation of the incoming photon, and such that the edge of the target facing the split unit is parallel to the $x=0$ plane.
The two wave packets have the form
\begin{equation}
\label{eqn: SMWavePack}
\ket{\gamma_j}=\frac{1}{\sqrt{\mathcal{A}}}\int dk \, c(\omega_k-\omega_0) e^{-ikx_j} \ket{1_{\mathbf{k}}}
\end{equation}
where $\ket{1_\mathbf{k}}$ are single photon states with momentum parallel to $x$, $c(\omega_k-\omega_0)$ is a function centered at $\omega_0$ with bandwidth $\Delta \omega \ll \omega_0$ and has the dimension of the square root of a length, $x_j$ is the distance the wave packet must travel to reach the target and $\mathcal{A}$ is the transverse area of the wave packet.

\subsection{The scattering state}
In what follows the international system of units is adopted. The full dynamics of the global system is generated by $\hat{H}_0$, which is the sum of the free Hamiltonians for the radiation and the target, plus an interaction term $\hat{H}_I$, which, assuming that the radiation is non-resonant with the target's internal structure~\cite{HAURIEGE}, can be written as
\begin{equation}
\label{eqn: SMIntHam}
\hat{H}_I= \frac{\hbar r_e c^2 }{4 \pi} \int d^3 k' d^3 k'' \frac{\hat{a}_{\mathbf{k}'}^\dagger \hat{a}_{\mathbf{k}''}}{\sqrt{\omega_{k'} \omega_{k''}}} \int_V d^3r \hat{\rho}(\mathbf{r}) e^{i (\mathbf{k}'' - \mathbf{k}')\cdot \mathbf{r}}\,.
\end{equation}
Here, $c$ is the speed of light in vacuum and $r_e \simeq 2.8 \times 10^{-15}$~m is the classical radius of the electron. $\hat{a}_{\mathbf{k}}$ and its conjugate are the photon destruction and creation operators, $\hat{\rho}(\mathbf{r})$ is the density of scatterers operator at point $\mathbf{r}$ and $V$ the volume of the target.
Denoting  the time evolution operator associated to $\hat{H}_0$ as $\hat{U}_0(t)$, the full time evolution operator can be evaluated perturbatively to give~\cite{MESSIAH}
\begin{equation}
\label{eqn: SMFullEvolution}
\hat{U}(t) \simeq \hat{U}_0 -\frac{i}{\hbar} \hat{U}_0(t)\int_0 ^t dt' \hat{U}_0^ \dagger (t') \hat{H}_I \hat{U}_0(t') dt'\,.
\end{equation}
The evolved state obtained by applying (\ref{eqn: SMFullEvolution}) to the initial state (\ref{eqn: SMInitState}) will contain a non-interacting contribution, due to the zero-th order term in the perturbative expansion of $\hat{U}(t)$, and a contribution involving interactions in which we are interested. By the linearity of $\hat{U}(t)$ this term is the sum of two contributions, corresponding to the two possible scattering channels for the photon ($j=1,2$)
\begin{equation}
\ket{\delta \Psi_j} = - \frac{i}{\hbar} \hat{U}_0(t)\int_0 ^t dt' \hat{U}_0^ \dagger (t') H_I \hat{U}_0(t')\, dt'\ket{\gamma_j}\ket{\psi}.
\end{equation}
Substituting (\ref{eqn: SMWavePack}) and (\ref{eqn: SMIntHam}) into this expression leads to
\begin{equation}
\label{eqn: SMEvolvedState}
\ket{\delta \Psi_j} =-i \frac{r_e c^2}{4 \pi \sqrt{\mathcal{A}}}  \hat{U}_0(t) \int_{0} ^t dt' \int_V d^3r \int d^3k' \int dk \frac{e^{-i(\mathbf{k}'\cdot\mathbf{r}-\omega_{k'} t')}}{\sqrt{\omega_{k'}}} \frac{e^{i(kx-\omega_k t')}}{\sqrt{\omega_k}} e^{-ikx_j} c(\omega_k-\omega_0) \hat{\rho}(\mathbf{r},t')\ket{\psi}\ket{1_{\mathbf{k}'}}\,.
\end{equation}
Here, the time-integration interval is considered large compared to the other timescales involved in the problem. Then, some of the integrals appearing can be approximated by the Fourier transforms of their respective integrands.

\subsection{Simplified case}
The probability of detecting a scattered photon at position $\mathbf{R}$ at time $t$ is \cite{SCULLY}
\begin{equation}
\begin{split}
\big(\bra{\delta \Psi_1}+\bra{\delta \Psi_2}\big)\hat{E}^{(+)}(\mathbf{R})&\hat{E}^{(-)}(\mathbf{R}) \big( \ket{\delta \Psi_1}+\ket{\delta \Psi_2} \big)=\\
=\big(\bra{\delta \Psi_1}+\bra{\delta \Psi_2}\big)\hat{E}^{(+)}(\mathbf{R})\hat{U}_0(t) & \hat{U}_0 ^\dagger (t)  \hat{E}^{(-)}(\mathbf{R}) \big( \ket{\delta \Psi_1}+\ket{\delta \Psi_2} \big)\,.
\end{split}
\end{equation}
Here, $\hat{E}^{(\pm)}(\mathbf{R})$ is the positive/negative frequency part of the electric field operator, corresponding to the destruction or creation of a photon at position $\mathbf{R}$. Since $\ket{\delta \Psi_j}$ are single photon states, the application of $\hat{E}^{(-)}(\mathbf{R})$ induces transitions to the electromagnetic vacuum  state $|0\rangle$~\cite{SCULLY}. Thus, we can insert an identity relation in the middle of the scalar product to give
\begin{equation}
\big(\bra{\delta \Psi_1}+\bra{\delta \Psi_2}\big)\hat{E}^{(+)}(\mathbf{R})\hat{U}_0(t) \ket{0}\bra{0} \hat{U}_0 ^\dagger (t)  \hat{E}^{(-)}(\mathbf{R}) \big( \ket{\delta \Psi_1}+\ket{\delta \Psi_2} \big)\,.
\end{equation}
It follows that the detection amplitude is given by $\bra{0} \hat{U}_0 ^\dagger (t)  \hat{E}^{(-)}(\mathbf{R}) \big( \ket{\delta \Psi_1}+\ket{\delta \Psi_2} \big)$ and in this case, it is given by the sum of two terms originating from the two different scattering channels. With the explicit form of the electric field operator,
\begin{equation}
\hat{E}^{(-)}(\mathbf{R})=i \sqrt{\frac{\hbar}{2 \epsilon_0 (2\pi)^{3}}}\int d^3 q \sqrt{\omega_q} a_{\mathbf{q}}e^{i\mathbf{q}\cdot\mathbf{R}}\,,
\end{equation}
($\epsilon_0$ being the vacuum permittivity) and using the explicit form of $\ket{\delta \Psi_j}$, the detection amplitude for the $j$-th channel becomes
\begin{equation}
\label{eqn: SMRoughDetAmpl}
\begin{split}
\bra{0} \hat{U}_0 ^\dagger (t)  \hat{E}^{(-)}(\mathbf{R}) \ket{\delta \Psi_j} = & -\frac{r_e c^2}{4\pi} \sqrt{\frac{\hbar}{2 (2\pi)^3 \mathcal{A} \epsilon_0} } \times\\
\times \int_{0} ^t dt' \int_V d^3r 
\hat{\rho}(\mathbf{r},t')\ket{\psi} 
\int d^3q e^{i [ \mathbf{q}\cdot (\mathbf{R}-\mathbf{r}) - \omega_q (t-t')]}
&\int dk e^{-i[\omega_k t' -k(x+x_j)]} \frac{c(\omega_k-\omega_0)}{\sqrt{\omega_k}}\,.
\end{split}
\end{equation}
Because of the properties of $c(\omega_k-\omega_0)$, we can approximate
\begin{equation}
\int dk e^{-i[\omega_k t' -k(x+x_j)]} \frac{c(\omega_k-\omega_0)}{\sqrt{\omega_k}} \simeq \frac{1}{\sqrt{\omega_0}} e^{-i \omega_0 \big( t'-\frac{x+x_j}{c} \big)} f\bigg( t'-\frac{x+x_j}{c} \bigg)\,,
\end{equation}
where $f$ is the envelope function of the photon wave packet and has a temporal extension $\sim 1/\Delta \omega$ and the dimension of an inverse square root of a length. The finite duration of the wave packet reduces the time-integration interval to the time the pulse needs to cross the target, which is of order $1/\Delta \omega +L/c$, with $L$ the longitudinal size of the target. Assuming that the dynamics of the scatterers in the target has slower timescale than the crossing time, $\hat{\rho}$ can be considered constant at the instant of arrival of the photon wave packet $t_j \equiv x_j/c$ and brought out of the time integral. The detection amplitude  then becomes
\begin{align}
-\frac{r_e c^2}{4\pi} \sqrt{\frac{\hbar}{2 (2\pi)^3 \omega_0 \mathcal{A} \epsilon_0} } \int_V d^3 & r 
\hat{\rho}(\mathbf{r},t_j)\ket{\psi}
\int d^3q e^{i [ \mathbf{q} \cdot (\mathbf{R}-\mathbf{r}) - \omega_q t]} \int_{0} ^t dt' f \bigg( t'-t_j-\frac{x}{c} \bigg) e^{i \omega_q t'} e^{-i \omega_0 \big( t'-t_j- \frac{x}{c} \big)} =
\\ \nonumber
 = -\frac{r_e c^2}{2} \sqrt{\frac{\hbar}{2 (2\pi)^3 \omega_0 \mathcal{A} \epsilon_0} } &\int_V d^3r 
\hat{\rho}(\mathbf{r},t_j)\ket{\psi}
\int d q \, d\theta \, q ^2 \sin\theta e^{i \omega_q \frac{|\mathbf{R}-\mathbf{r}|}{c} \cos\theta}e^{-i \omega_q \big( t - t_j - \frac{x}{c} \big)} c(\omega_q-\omega_0)\,.
\end{align}
Upon integration over $\theta$, taking into account that the distance of the detection point is much larger than the size of the target, the scattering amplitude becomes
\begin{equation}
-\frac{r_e}{2} \sqrt{\frac{\hbar \omega_0}{2 (2\pi)^3 \mathcal{A} \epsilon_0} }\frac{e^{i \omega_0 ( R/c - t)}-e^{-i \omega_0 ( R/c + t)}}{R} e^{i \omega_0 t_j} \int_V d^3r  e^{-i(\tilde{\mathbf{k}}_0-\mathbf{k}_0)\cdot \mathbf{r}} f\bigg( t- t_j- \frac{|\mathbf{R}-\mathbf{r}|+x}{c} \bigg)  \hat{\rho}(\mathbf{r},t_j)\ket{\psi}\,,
\end{equation}
where $\tilde{\mathbf{k}}_0 \equiv \mathbf{R} \omega_0/(cR) $ and $\mathbf{k}_0 \equiv \mathbf{x} \omega_0/(cx)$. The last line is a superposition of outgoing and ingoing spherical waves centered at the target, of which the ingoing does not correspond to the boundary conditions of interest here and therefore is dropped.
Supposing moreover that the envelope does not vary significantly over the size of the target, i.e., $L \ll c/ \Delta\omega $, and neglecting the propagation time to the detector, one finds the final form of the detection amplitude
\begin{equation}
\label{eqn: SMDetAmpl}
\bra{0} \hat{U}_0 ^\dagger (t)  \hat{E}^{(-)}(\mathbf{R}) \ket{\delta \Psi_j}=-\frac{r_e}{2} \sqrt{\frac{\hbar \omega_0}{2 (2\pi)^3 \mathcal{A} \epsilon_0} }\frac{e^{i \omega_0 ( R/c - t)}}{R} e^{i \omega_0 t_j} f ( t- t_j ) \int_V d^3r  e^{-i(\tilde{\mathbf{k}}_0-\mathbf{k}_0)\cdot \mathbf{r}} \hat{\rho}(\mathbf{r},t_j)\ket{\psi}\,.
\end{equation}
After the interaction with the target, the overlap unit creates a contribution in which the two scattering pathways temporally overlap, by delaying the advanced one.

\subsection{Realization with M\"ossbauer foils}
In the original proposal~\cite{BARON}, the split and overlap units consist of two identical M\"ossbauer foils containing $^{57}$Fe. When a pulse impinges on a M\"ossbauer foil, the component of the pulse resonant with the M\"ossbauer transition at frequency $\omega_0\simeq 14.4$~keV is scattered on a time scale $T\simeq 141$~ns.
One of the two foils used in the original arrangement is slightly detuned from the other by moving it at a constant velocity, this causing a Doppler shift $\Omega \gg T^{-1}$ of its transition energy. This detuning has the additional advantage that no pathway exists in which a single photon scatters in both foils.  In the following, we assume that the split unit is subject to the Doppler shift.
Thus, the initial state of the photon is a superposition of a state of the kind (\ref{eqn: SMWavePack}), that is a copy of the temporally short incoming wave packet, and a temporally long wave packet with an approximately Lorentzian spectral shape $\mathscr{L}(\omega_k-\omega_0-\Omega)$. The Lorentzian spectral shape corresponds to an approximately exponential decay in the time domain, which starts immediately after the excitation, such that  $x_1=x_2 \equiv x_0$.
The first component of the initial photon state gives a contribution to the detection amplitude similar to (\ref{eqn: SMDetAmpl}). However, the second foil acts on it by scattering its component at frequency $\omega_0$, such that the envelope $f$ becomes an exponentially decaying function
\begin{equation}
f(\xi) \longrightarrow h(\xi) \equiv \Theta(\xi)e^{-\frac{\xi}{T}}
\end{equation}
with $\Theta$ the Heaviside step function.
The scattering of the Lorentzian part of the initial photon state needs a different treatment because the characteristic time $T$ of its exponentially decaying envelope is comparable with timescale of the internal dynamics of the target.
This fact does not allow us to assume that the scatterer's density operator is constant during the crossing time of the wave packet. In order to take into account the time dependency of $\hat{\rho}$ in formula (\ref{eqn: SMRoughDetAmpl}) we decompose it into the energy eigenvectors of the target, 
\begin{equation}
\hat{\rho}(\mathbf{r},t')=\sum_{m,n} e^{i\omega_{mn}t'} \bra{m} \hat{\rho}(\mathbf{r}) \ket{n} \ket{n}\bra{m}\,,
\end{equation}
where $\omega_{mn}$ are the characteristic frequencies of the target's internal dynamics. Since we assume that  they are non-resonant with the radiation, they are not within the support of the Lorentzian wave packet.
As a consequence, the detection amplitude due to the second component of the initial photon state is
\begin{align}
-\frac{r_e c^2}{4\pi} \sqrt{\frac{\hbar}{2 (2\pi)^3 \omega_0 \mathcal{A} \epsilon_0} } \sum_{m,n} & \int_V d^3r 
\bra{m}\hat{\rho}(\mathbf{r})\ket{n}\inner{n}{\psi} \ket{m} \times \\
\times \int d^3q e^{i [ \mathbf{q}\cdot (\mathbf{R}-\mathbf{r}) - \omega_q t]}
\int_{0} ^t dt' e^{i(\omega_q+\omega_{mn})t'} & \int dk e^{-i[\omega_k t' -k(x+x_0)]} \frac{\mathscr{L}(\omega_k-\omega_0-\Omega)}{\sqrt{\omega_k}}
\end{align}
The explicit calculation of the integrals over time and momenta transform the last expression into
\begin{equation}
-\frac{r_e}{2} \sqrt{\frac{\hbar \omega_0}{2 (2\pi)^3 \mathcal{A} \epsilon_0} }\frac{1}{R}\int_V d^3 r \hat{\rho}\bigg( \mathbf{r}, t- \frac{|\mathbf{R}-\mathbf{r}|}{c} \bigg) \ket{\psi} h\bigg( t- \frac{|\mathbf{R}-\mathbf{r}|+x+x_0}{c} \bigg) e^{i (\omega_0+\Omega)\big( t- \frac{|\mathbf{R}-\mathbf{r}|+x+x_0}{c} \big)}\,.
\end{equation}
Assuming that $T \gg L/c$, defining the wave vectors $\tilde{\mathbf{k}}_0 \equiv (\omega_0+\Omega)\mathbf{R}/(cR)\simeq \omega_0\mathbf{R}/(cR)$ and $\mathbf{k}_0 \equiv (\omega_0+\Omega)\mathbf{x}/(cx)\simeq \omega_0\mathbf{x}/(cx)$ the detection amplitude in the second channel is
\begin{equation}
\begin{split}
-\frac{r_e}{2} \sqrt{\frac{\hbar \omega_0}{2 (2\pi)^3 \mathcal{A} \epsilon_0} } \frac{e^{i \omega_0(R/c - t)}}{R} e^{i \Omega (t-\frac{x_0}{c})}  h\bigg( t- \frac{R+x_0}{c} \bigg) \int_V d^3 r \hat{\rho}\bigg( \mathbf{r}, t- \frac{R}{c} \bigg) \ket{\psi} e^{-i (\tilde{\mathbf{k}}_0-\mathbf{k})\cdot \mathbf{r}}
\end{split}
\end{equation}
It can be assumed that the second foil has no effect on the scattered photon because, due to the Doppler shift, the spectrum of the latter is far from the resonance of the former. In addition the factor $\Omega x_0/c$  usually is very small giving no relevant phase contributions. The total amplitude for detection after the overlap unit is then
\begin{align}
-\frac{r_e}{2} \sqrt{\frac{\hbar \omega_0}{2 (2\pi)^3 \mathcal{A} \epsilon_0} } \frac{e^{i \omega_0(R/c - t)}}{R} & e^{i \omega_0 \frac{x_0}{c}}  h\bigg( t- \frac{R+x_0}{c} \bigg) \times\\
\times \bigg(\int_V d^3 r \hat{\rho} ( \mathbf{r}, t_0 ) \ket{\psi} e^{-i (\tilde{\mathbf{k}}_0-\mathbf{k})\cdot \mathbf{r}}
+ e^{i \Omega t } \int_V & d^3 r \hat{\rho}\bigg( \mathbf{r}, t- \frac{R}{c} \bigg) \ket{\psi} e^{-i (\tilde{\mathbf{k}}_0-\mathbf{k})\cdot \mathbf{r}} \bigg)
\end{align}
which is the quantum correspondent of the result found in \cite{BARON}. Note that the phase difference in this case is given by the Doppler-shift factor $\Omega t$ and that an additional phase shift $\phi$ must be added to it if the phase control technique developed in reference \citep{Heeg2017} is used.

\section{Calculation of DCF for one particle in a Double Well potential}
When the particles in the target system can only occupy discrete positions, the integral defining the DCF at the separation $\mathbf{r}$ [Eq. (2) in the main text] reduces to a discrete sum over all the possible pairs of points with mutual distance $\mathbf{r}$,
\begin{equation}
G_{qu}(\mathbf{r},t_1,t_2)=\sum_{\mathbf{r}'_\mathbf{n}} \text{Tr} \big[ \mu \hat{\rho}_\mathbf{n} (t_1) \hat{\rho}_{\mathbf{n}+\mathbf{r}}(t_2) \big]=\sum_{\mathbf{r}'_\mathbf{n}} \text{Tr} \big[ \mu(t_1) \hat{\rho}_\mathbf{n} \hat{\rho}_{\mathbf{n}+\mathbf{r}}(t_2-t_1) \big]\,.
\end{equation}
The ISF then is given by a discrete Fourier transform
\begin{equation}
S_{qu}(\mathbf{p},t_1,t_2)=\sum_{\mathbf{r}}G_{qu}(\mathbf{r},t_1,t_2) e^{i \mathbf{p}\cdot\mathbf{r}}\,.
\end{equation}
In our case, the particle can only occupy the two minima of the double-well potential labeled by $L,R$ whose distance is $\mathbf{d}$. As a consequence, the DCF has only three values in correspondence of the three distances $\mathbf{r}=0,\pm \mathbf{d}$, given by
\begin{align}
G_{qu}(\mathbf{d},t_1,t_2)&=\text{Tr} \big[ \mu(t_1) \hat{\rho}_L \hat{U}^{\dagger}(t_2-t_1)\hat{\rho}_R \hat{U}(t_2-t_1)\big]\,,\\
G_{qu}(-\mathbf{d},t_1,t_2)&=\text{Tr} \big[ \mu(t_1) \hat{\rho}_R \hat{U}^{\dagger}(t_2-t_1)\hat{\rho}_L \hat{U}(t_2-t_1)\big]\,,\\
G_{qu}(0,t_1,t_2)&=\text{Tr} \big[ \mu(t_1) \hat{\rho}_L \hat{U}^{\dagger}(t_2-t_1)\hat{\rho}_L \hat{U}(t_2-t_1)\big] + \nonumber \\
&\qquad +\text{Tr} \big[ \mu(t_1) \hat{\rho}_R \hat{U}^{\dagger}(t_2-t_1)\hat{\rho}_R \hat{U}(t_2-t_1)\big]\,.
\end{align}
As only one particle is considered, the explicit form of the density operators at the two position $L,R$ are simply
$\hat{\rho}_L = \ket{L}\bra{L}$ and $\hat{\rho}_R = \ket{R}\bra{R}$.
The Hamiltonian for the single particle is
\begin{equation}
\hat{H}=-\hbar \frac{\omega_{h}}{2} \big( \ket{L}\bra{R}+\ket{R}\bra{L} \big)\,,
\end{equation}
and the time evolution operator can be calculated exactly to give
\begin{equation}
\label{Uexplicit}
\hat{U}(t)=\cos\bigg( \frac{\omega_{h}}{2}t \bigg)+i \big( \ket{L}\bra{R}+\ket{R}\bra{L} \big) \sin\bigg( \frac{\omega_{h}}{2}t \bigg)\,.
\end{equation}
The expression of the evolved density matrix is kept implicit in the calculation as we wanted to point out the role of coherences at the time at which the first variable is considered, but its explicit time dependency follows in a straightforward way from (\ref{Uexplicit}).

\end{document}